\documentstyle[12pt]{article}

\newcommand{\resetcounter}{\setcounter{equation}{0}}     

\begin{document}

\thispagestyle{empty}
\begin{titlepage}
\begin{flushright}
HUB-IEP-95/18 \\
hep-th/9509146 \\
September 1995
\end{flushright}
\vspace{0.3cm}
\begin{center}
\Large \bf String-dual $F_1$-Function
           \\ in the Three Parameter Model
\end{center}
\vspace{0.5cm}
\begin{center}
Gottfried \ Curio$^{\hbox{\footnotesize{1,2}}}$  \\
{\sl Institut f\"ur Physik, Humboldt--Universit\"at,\\
 Invalidenstrasse 110, D--10115 Berlin, Germany}
\end{center}
\vspace{0.6cm}

\begin{abstract}
\noindent
Computation of the holomorphic $F_1$-function describing one-loop
gravitational couplings of
vectormultiplets is shown to confirm
string-string duality
of the proposed dual pair consisting of the heterotic string
on $K3\times T^2$ with gauge group $U(1)^4$ and the type IIA
string on the Calabi-Yau $WP^4_{1,1,2,8,12}(24)$.
\end{abstract}

\vspace{0.3cm}
\footnotetext[1]{E-MAIL: curio@qft2.physik.hu-berlin.de}
\footnotetext[2]{Supported by Deutsche Forschungsgemeinschaft}
\vfill
\end{titlepage}


\setcounter{page}{1}

\resetcounter
In the course of establishing finer tests of string-string duality
[\ref{D}],[\ref{W}],[\ref{FHSV}],[\ref{KV}],[\ref{KLM}],[\ref{KKLMV}],
[\ref{AGNT}],[\ref{AP}]  agreement was found to some orders [\ref{KLTh}]
 resp. analytically [\ref{C}]  when comparing the holomorphic
 $F_1$-functions - describing one-loop gravitational couplings of
 vectormultiplets - for the heterotic string on $K3\times T^2$
with gauge group $U(1)^3$ and the type IIA string on $WP^4_{1,1,2,2,6}(12)$
. Here I extend this result to the other main example
 of [\ref{KV}],[\ref{KLM}],[\ref{KKLMV}]: the heterotic string
on $K3\times T^2$ with gauge group $U(1)^4$ (now with
 both toroidal moduli T and U free and no longer restricted to the diagonal)
 and the type IIA string on $WP^4_{1,1,2,8,12}(24)$ [\ref{HKThY}].\\

 The defining polynomial of $WP^4_{1,1,2,8,12}(24)^{-480}_{3,243}$ is

\begin{eqnarray*}
p=z_1^{24}+z_2^{24}+z_3^{12}+z_4^3+z_5^2+a_0z_1z_2z_3z_4z_5+a_1(z_1z_2z_3)^4
+a_2(z_1z_2)^{12} \; ,
\end{eqnarray*}


which shows already the deformations of the mirror Calabi-Yau
with uniformizing variables at large complex
structure\footnote{I  use also the parameters
 $a_0=-12\psi_0,\, a_1=-2\psi_1,\, a_2=-\psi_2$
and $\bar{x} =\frac{\alpha}{4}x, \; \bar{y} =4y, \;
 \bar{z} =4z $, where $\alpha = j(i) = 1728$.}

\begin{eqnarray*}
x=-\frac{2}{\alpha^2}\psi_0^{-6}\psi_1, \;
y=\psi_2^{-2}, \;
z=-\frac{1}{4}\psi_1^{-2}\psi_2 \; .
\end{eqnarray*}

For later use let us also record the inverse relations

\begin{eqnarray*}
\psi_0 &\sim& \bar{y}^{-\frac{1}{24}}(\bar{x}^2\bar{z})^{-\frac{1}{12}}  \\
\psi_1 &\sim& (\bar{y}^{-\frac{1}{2}}\bar{z}^{-1})^{\frac{1}{2}}  \\
\psi_2 &\sim& \bar{y}^{-\frac{1}{2}}.
\end{eqnarray*}

 Like [\ref{CdlOFKM}] I will use discriminant factors shifted
 compared to [\ref{KLM}],[\ref{HKThY}](we are  in the
 limit $\bar{y}\rightarrow 0$)

\begin{eqnarray*}
\tilde{\Delta}_1 &=& \bar{y}^{-1}(\frac{1-\bar{z}}{\bar{z}})^2  \\
\tilde{\Delta}_2 &=& \bar{y}^{-1}(\bar{x}^2\bar{z})^
{-2}((1-\bar{x})^2-\bar{x}^2\bar{z})^2  \\
\tilde{\Delta}_3 &=& \frac{1-\bar{y}}{\bar{y}}\approx\bar{y}^{-1} \; .
\end{eqnarray*}


 In the  limit $\bar{y}\rightarrow 0$ the mirror map is given by\footnote{Here
$q_1:=q_T, \; q_3:=q_U/q_T; \; j:=j(iT), \; k:=j(iU)$.} [\ref{KLM}], [\ref{LY}]

\begin{eqnarray*}
\bar{x} &=& \frac{\alpha}{2}\frac{j+k-\alpha}{jk+\sqrt{j(j-\alpha)}
                     \sqrt{k(k-\alpha)}}
                =q_1+\sum_{m+n>1}a_{mn}q_1^mq_3^n  \\
\bar{y} &=& q_2f_y(q_1,q_3)+O(q_2^2)\qquad\qquad
                =q_2\sum_{m+n\ge 1}c_{mn}q_1^mq_3^n+O(q_2^2) \\
\bar{z} &=& (\frac{\alpha}{2})^2\frac{1}{jk\bar{x}^2}\qquad\qquad\qquad\qquad
               =q_3+\sum_{m+n>1}b_{mn}q_1^mq_3^n \; .
\end{eqnarray*}

One has $t_1=t_T=iT, \; t_2=4\pi iS, \; t_3=t_U-t_T,\;t_U=iU$;
 so $q_2=e^{-8\pi^2S}$, where S is the tree-level dilaton of the
 heterotic string, and $\bar{y}=e^{-8\pi^2S^{inv}}$ with the modular
 invariant dilaton [\ref{dWKLL}], [\ref{KLTh}].  \\
 Now on the heterotic side at weak coupling one has
[\ref{L.-C.LM}],[\ref{dWKLL}]

\begin{eqnarray*}
F_1^{het} \sim \log[\bar{y}^{-\alpha}(j-k)^{\beta}(\eta^{-2}
(t_T)\eta^{-2}(t_U))^{\gamma}]
\end{eqnarray*}

with $\alpha=2$ and

\begin{eqnarray*}
\alpha &=& \frac{24}{4}\beta  \\
\gamma &=& \frac{b_{grav}}{4}\beta   \; ,
\end{eqnarray*}

where $528=b_{grav}=48-\chi$  [\ref{KLTh}] and
$528=22\cdot24=44\cdot12$ with $\gamma=44$ and $\beta=\frac{1}{3}$.  \\
Namely  in

\begin{eqnarray*}
F_1^{het} &=& 24S^{inv}+\frac{1}{4\pi^2}\cdot2\cdot\log(j-k)-
\frac{b_{grav}}{4\pi^2}\cdot\frac{1}{2}\cdot\log\eta^2(iT)\eta^2(iU)\\
          &=&\frac{6}{4\pi^2}\log[\bar{y}^{-2}(j-k)^{\frac{2}{6}}
(\eta^{-2}(iT)\eta^{-2}(iU))^{\frac{b_{grav}/2}{6}}]
\end{eqnarray*}

one has two additional factors of 2 relative to the situation
in the diagonal T=U-model\footnote{Note that in the process of
restriction to the diagonal the possible symmetry enhancement
 at $T=U=\rho$ is lost as after the step $U(1)^2\rightarrow
 SU(2)\times U(1)$ is done the SU(3) is out of reach. So the
 results in the two parameter model are not a simple truncation
 of the case considered here.} of [\ref{KLTh}],[\ref{C}], where

\begin{eqnarray*}
F_1^{het} &=& 24S^{inv}+\frac{1}{4\pi^2}\log(j-\alpha)-
\frac{b_{grav}}{4\pi^2}\log\eta^2(iT)\\
          &=&\frac{6}{4\pi^2}\log[\bar{y}^{-2}(j-\alpha)
^{\frac{1}{6}}\eta^{-2}(iT)^{\frac{b_{grav}}{6}}] \; .
\end{eqnarray*}

The first factor of 2 comes from the different (generic) pole order
 of $\log(j-k)$ in $\log(T-U)$ compared to that of $\log(j-\alpha)$
in $\log(T-1)$, the second one from the fact that the one-loop quantity
 $h^{(1)}$ (cf.[\ref{dWKLL}],[\ref{KLTh}]) has weights (-2,-2) in T and U
 up to a quadratic polynomial, so one needs for modular invariance
 $\frac{b_{grav}}{16\pi^2}\log(\eta^2(iT)\eta^2(iU))^2$ with the
 weight (2,2) quantity inside the logarithm.\\

 On the Calabi-Yau side one gets from the holomorphic anomaly of $F_1$
 [\ref{BCOV}] that up to an additive constant (cf.[\ref{KLTh}]
, [\ref{CdlOFKM}], [\ref{C}] )

\begin{eqnarray*}
F_1^{II} = \log[(\frac{24\psi_0}{\omega_0})^{3+3-\frac{\chi}{12}}
\frac{\partial(\psi_0,\psi_1,\psi_2)}{\partial(T,S,U)}f]
\end{eqnarray*}

with a holomorphic function {\em f} \,  and the fundamental period
$\omega_0=(E_4(T)E_4(U))^{\frac{1}{4}}$. Note that $\psi_0/\omega_0$
is regular and $\neq0$ at $\psi_0\approx0$ since one has $\psi_0
\sim (jk)^{12} \sim (E_4(T)E_4(U))^{\frac{1}{4}}$  as far as the
behaviour relevant here is concerned. So {\em f}\,   is of the form
$f=\tilde{\Delta}_1^a \tilde{\Delta}_2^b \tilde{\Delta}_3^d \psi_0^c $.\\

At first sight the form of these factors of {\em f}\,  looks as if here
additional enhancements  besides the diagonal one ( and its further
ramifications ) known from the heterotic side could occur, but in view
of the  crucial relation\footnote{Cf. the direct computation in the Appendix;
 a conceptual argument can also be given (cf. also [\ref{LY}]).}
 $\tilde{\Delta}_1\tilde{\Delta}_2 \sim (j-k)^4$ \,   we just arrive at
 the prediction that one must have $a=b$ .\\

Now one has $\frac{\partial(\bar{x},\bar{y},\bar{z})}
{\partial(\psi_0,\psi_1,\psi_2)} \sim \psi_0^{-7}\psi_1^{-2}\psi_2^{-2}$
and $\frac{\partial(\bar{x},\bar{y},\bar{z})}{\partial(T,S,U)} \sim
\bar{y}(\bar{x}_U\bar{z}_T-\bar{x}_T\bar{z}_U)$, where  $\bar{x}_T=-j_T
\cdot\frac{\bar{x}(1-\bar{x})}{j+k-\alpha}\cdot\frac{\sqrt{k(k-\alpha)}}
{\sqrt{j(j-\alpha)}}\sim-j_T\cdot\frac{1}{jk}\cdot\frac{\sqrt{k(k-\alpha)}}
{\sqrt{j(j-\alpha)}}$ and $\bar{z}_T=-\frac{j_T}{j}\bar{z}-2\bar{z}
\frac{\bar{x}_T}{\bar{x}}$, so with the relation $j_T^2 \sim j(j-\alpha)E_4$
 it follows that

\begin{eqnarray*}
\frac{\partial(\psi_0,\psi_1,\psi_2)}{\partial(T,S,U)}
 &\sim& \psi_0^7\psi_1^2\psi_2^2 \bar{y} \frac{\bar{z}}{jk}
        j_Tk_U\frac{j-k}{\sqrt{k(k-\alpha)}\sqrt{j(j-\alpha)}} \\
 &\sim& \bar{y}^{-\frac{3}{4}}\frac{j-k}{\sqrt{jk}}\omega_0^2\psi_0 .
\end{eqnarray*}
 The behaviour of the jacobian near $\psi_0=0$ restricts c to 3
because (up to the $\bar{y}$-dependence ) the jacobian behaves there in
$\psi_0$
as $(j-k)(\psi_0^{12})^{-\frac{1}{2}}\psi_0^2\psi_0$.\\

 So one is led to
\begin{eqnarray*}
F_1^{II}&=&\log[\bar{y}^{-[2+(\frac{c+1+16}{24}+a+b+d)]}
(\omega_0^2)^{-22}(jk)^{\frac{40+c+1}{12}}(j-k)(\frac{1-\bar{z}}
{\bar{z}})^{2a}(\frac{(1-\bar{x})^2-\bar{x}^2\bar{z}}
{\bar{x}^2\bar{z}})^{2b}]\;.
\end{eqnarray*}

 To determine a, b, d let us at the large radius limit compare
this to the topological intersection numbers. In the two steps
 $U\rightarrow\infty$ ( $\Rightarrow\bar{z}\rightarrow0$ )
 and then $T\rightarrow\infty$  , one has
 (up to the $\bar{y}$-dependence and numerical factors)

\begin{eqnarray*}
 j-k&\rightarrow&k\qquad\qquad\;\;\;\;\rightarrow k \\
\bar{x}&\rightarrow&\frac{1}{j+\sqrt{j(j-\alpha)}}\rightarrow\frac{1}{j}\\
 \bar{z}&\rightarrow& \bar{z}\qquad\qquad\;\;\;\;\;\;\;\rightarrow
\frac{j^2}{jk}\\
\tilde{\Delta}_1&\rightarrow&\bar{z}^{-2}\qquad\quad\;\;\;\;\;\;\;
\rightarrow(\frac{k}{j})^2\\
\tilde{\Delta}_2&\rightarrow&\frac{(\bar{x}^{-1}-1)^4}{\bar{z}^2}
\;\;\;\;\;\;\rightarrow(jk)^2 \; ,
\end{eqnarray*}

so

\begin{eqnarray*}
F_1^{II}\rightarrow\log[\bar{y}^{-[2+(\frac{5}{6}+a+b+d)]}
(jk)^{\frac{44}{12}}k(\frac{k}{j})^{2a}(jk)^{2b}] \; ,
\end{eqnarray*}

which gives for the large radius limit

\begin{eqnarray*}
& &-\frac{2\pi i}{12}[(24+12(\frac{5}{6}+a+b+d))t_2+
(44+12(2b-2a))t_T+(44+12(1+2a+2b))t_U]\\
&=&-\frac{2\pi i}{12}[(24+12(\frac{5}{6}+a+b+d))t_2+
(2\cdot44+12(1+4b))t_1+(44+12(1+2a+2b))t_3] \; .
\end{eqnarray*}

Now compare (as [\ref{CdlOFKM}]) this with\footnote
{Note that in [\ref{HKThY}] (A.38) you have to read
$\frac{1}{4}<u,l^{(2)}+2l^{(3)}>$ in front of
$e_{\nu^*_6}$ and to overcross in the identification
 of $-e_{\nu^*_6}\,,-e_{\nu^*_7}$ and $h_D\,,h_E$.}

\begin{eqnarray*}
 -\frac{2\pi i}{12}[c_2\cdot J \,t_1\,+c_2\cdot
\frac{J-2D-E}{4}\,t_2\,+c_2\cdot\frac{J-E}{2}\,t_3]\;,
\end{eqnarray*}

where [\ref{HKThY}] J is the hyperplane class of weight 4
(cf.[\ref{BK}]), L the $K_3$-fibre belonging to the linear
system of degree one polynomials, D the elliptic ruled surface
of the resolved singular curve and E the Hirzebruch surface
of the resolution of the special point on the singular curve.
 Now(cf.[\ref{CdlOFKM}]) $c_2\cdot X=c_2(X)-X^3$, so with
$c_2\cdot L=24-0, \, c_2\cdot D=0-0, \, c_2\cdot E=4-8,$
one gets $a=b,\, a=-1/6,\, d=-1/2$, i.e. (using $j=E_4^3/\eta^{24}$)

\begin{eqnarray*}
F_1^{II}&=&\log[\bar{y}^{-2}(\omega_0^2)^{-22}
(jk)^{\frac{44}{12}}(j-k)^{1+4a}]\\
        &=&\log[\bar{y}^{-2}(j-k)^{1/3}(\eta^{-2}
(iT)\eta^{-2}(iU))^{44}]\,,
\end{eqnarray*}

which matches the actual heterotic values.\\
To get also their cohomological meaning just note
(besides the direct observation $\alpha=\chi_{K3}/12$)
that you get from the general ansatz in $\alpha,\beta,\gamma$
that\footnote{In the last transformation of the
following equation series I use that (because of the K3-fibration)
one has $\chi=(\chi_{P^1}-24)\chi_{K3}+24(0-1+\chi_{P^1}+1)$
so $-\frac{\chi}{12}=-4+2\chi_{K3}-2\chi_{P^1}$; here the +1
comes from the possibility of having $z_3=z_4=z_5=0$
and the $-1+\chi_{P^1}$ from the replacement of the special point
 of the elliptic curve by the base of its resolving Hirzebruch surface.}

\begin{eqnarray*}
(c+1-\frac{\chi}{12})(12\frac{\beta}{\gamma}+2)=12\beta
+2\gamma=c_2\cdot J=4\chi_{K3}+c_2(E)-E^3=
4\chi_{K3}+2\chi_{P^1}-4\chi_{P^1}\\=
2\cdot2(\chi_{K3}-\chi_{P^1})+2\chi_{P^1}=2(4-\frac{\chi}{12})+2\chi_{P^1}\;,
\end{eqnarray*}

so indeed $\frac{\gamma}{\beta}=\frac{b_{grav}}{4}$
(and $\frac{\alpha}{\beta}=\frac{\chi_{K3}}{12}\cdot
\frac{b_{grav}/4}{\gamma}=\frac{\chi_{K3}}{4}$ as
 $\gamma=c+1-\frac{\chi_{K3}}{12}=\frac{b_{grav}}{12}$).\\
This agrees with the meaning of the numbers on the heterotic side.


\hspace{0.5cm}

{\bf Acknowledgements:} I would like to thank  E. Derrick,
 G. Lopes-Cardoso , D. L\"ust for  discussions.

\hspace{0.5cm}

{\bf Appendix}\\\\

The relation $\sqrt{\tilde{\Delta}_1}\
sqrt{\tilde{\Delta}_2}\sim(j-k)^2$ is shown:

\begin{eqnarray*}
& &\frac{1-\bar{z}}{\bar{z}}\times
\frac{(1-\bar{x})^2-\bar{x}^2\bar{z}}{\bar{x}^2\bar{z}}\\
&\sim& jk(\frac{j+k-\alpha}{jk+\sqrt{j(j-\alpha)}\sqrt{k(k-\alpha)}})^2
\frac{jk(j+k-\alpha)^2-(jk+\sqrt{j(j-\alpha)}\sqrt{k(k-\alpha)})^2}
{jk(j+k-\alpha)^2} \times\\
& & jk[\frac{(jk+\sqrt{j(j-\alpha)}
\sqrt{k(k-\alpha)}-\frac{\alpha}{2}
(j+k-\alpha))^2}{(jk+\sqrt{j(j-\alpha)}
\sqrt{k(k-\alpha)})^2} - \frac{\alpha^2}{4jk}]\\
&=&jk(\frac{j+k-\alpha}{jk+\sqrt{j(j-\alpha)}
\sqrt{k(k-\alpha)}})^2
\frac{j^2+k^2-\alpha(j+k)-2\sqrt{j(j-\alpha)}
\sqrt{k(k-\alpha)}}{(j+k-\alpha)^2}\times\\
& &jk[\frac{\alpha}{2}\frac{jk+\sqrt{j(j-\alpha)}\sqrt{k(k-\alpha)}
-\frac{\alpha}{2}(j+k-\alpha)}{(jk+\sqrt{j(j-\alpha)}\sqrt{k(k-\alpha)})^2}
 \frac{2}{\alpha}(jk+\sqrt{j(j-\alpha)}\sqrt{k(k-\alpha)}\\
& &\,-\frac{\alpha}{2}(j+k-\alpha))- \frac{\alpha^2}{4jk}]\\
&=&jk(\frac{j+k-\alpha}{jk+\sqrt{j(j-\alpha)}\sqrt{k(k-\alpha)}})^2
 (\frac{\sqrt{j(j-\alpha)}-\sqrt{k(k-\alpha)}}{j+k-\alpha})^2 \times\\
& &jk[\frac{\alpha}{4jk} \frac{2}{\alpha} (jk+\sqrt{j(j-\alpha)}
\sqrt{k(k-\alpha)}-\frac{\alpha}{2}(j+k-\alpha)) -\frac{\alpha^2}{4jk}]\\
&=&jk(\frac{j+k-\alpha}{jk+\sqrt{j(j-\alpha)}\sqrt{k(k-\alpha)}})^2
 (\frac{j(j-\alpha)-k(k-\alpha)}{(j+k-\alpha)(\sqrt{j(j-\alpha)}+
\sqrt{k(k-\alpha)})})^2 \times\\
& &\frac{1}{4}[2jk+2\sqrt{j(j-\alpha)}\sqrt{k(k-\alpha)}-\alpha(j+k)]\\
&\sim&jk(\frac{j+k-\alpha}{jk+\sqrt{j(j-\alpha)}\sqrt{k(k-\alpha)}})^2
(\frac{(j-k)(j+k-\alpha)}{(j+k-\alpha)
(\sqrt{j(j-\alpha)}+\sqrt{k(k-\alpha)})})^2 \times\\
& &[2jk+2\sqrt{j(j-\alpha)}\sqrt{k(k-\alpha)}-\alpha(j+k)]\\
&=&(j-k)^2jk(\frac{j+k-\alpha}{(jk+\sqrt{j(j-\alpha)}
\sqrt{k(k-\alpha)})(\sqrt{j(j-\alpha)}+\sqrt{k(k-\alpha)})})^2\times\\
& &[2jk+2\sqrt{j(j-\alpha)}\sqrt{k(k-\alpha)}-\alpha(j+k)]\\
&=&(j-k)^2jk(\frac{j+k-\alpha}{(jk+k(k-\alpha))\sqrt{j(j-\alpha)}+
(jk+j(j-\alpha))\sqrt{k(k-\alpha)}})^2\times\\
& &[2jk+2\sqrt{j(j-\alpha)}\sqrt{k(k-\alpha)}-\alpha(j+k)]\\
&=&(j-k)^2jk(\frac{1}{k\sqrt{j(j-\alpha)}+j\sqrt{k(k-\alpha)}})^2
\times[2jk+2\sqrt{j(j-\alpha)}\sqrt{k(k-\alpha)}-\alpha(j+k)]\\
&=&(j-k)^2\frac{1}{k(j-\alpha)+j(k-\alpha)+2\sqrt{j(j-\alpha)}
\sqrt{k(k-\alpha)}}\times\\
& &[2jk+2\sqrt{j(j-\alpha)}\sqrt{k(k-\alpha)}-\alpha(j+k)]\\
&=&(j-k)^2 \; .
\end{eqnarray*}

Here in going to the fourth line the relation $\frac{\bar{x}
(1-\bar{x})}{j+k-\alpha}=\frac{\alpha}{4jk}$ was used.

%
%

\section*{References}
\begin{enumerate}
\item
\label{D}
M. Duff, Nucl. Phys. {\bf B442} (1995) 47, hep-th/9501030.

\item
\label{W}
E. Witten, Nucl. Phys. {\bf B443} (1995) 85, hep-th/9503124.

\item
\label{FHSV}
S. Ferrara, J. Harvey, A. Strominger, C. Vafa, hep-th/9505162.

\item
\label{KV}
S. Kachru, C. Vafa, hep-th/9505105.

\item
\label{KLM}
A. Klemm, W. Lerche, P. Mayr, hep-th/9506112.

\item
\label{KKLMV}
S. Kachru, A. Klemm, W. Lerche, P. Mayr, C. Vafa, hep-th/9508155.

\item
\label{AGNT}
I. Antoniadis, E. Gava, K.S. Narain, T.R. Taylor, hep-th/9507115.

\item
\label{AP}
I. Antoniadis, H. Partouche, hep-th/9509009.

\item
\label{L.-C.LM}
G. Lopes-Cardoso, D. L\"ust, T. Mohaupt, Nucl. Phys. {\bf B450} (1995) 115.

\item
\label{dWKLL}
B. de Wit, V. Kaplunovsky, J. Louis, D. L\"ust, hep-th/9504006.

\item
\label{KLTh}
V. Kaplunovsky, J. Louis, S. Theisen, hep-th/9506110.

\item
\label{BCOV}
M. Bershadsky, S. Cecotti, H. Ooguri, C. Vafa, Nucl. Phys.
 {\bf B405} (1993) 279; Comm. Math. Phys. {\bf 165} (1994) 31.

\item
\label{CdlOFKM}
P. Candelas, X.C. de la Ossa, A. Font, S. Katz, D. Morrison,
Nucl. Phys. {\bf B416} (1994) 481.

\item
\label{C}
G.Curio, hep-th/9509042.

\item
\label{HKThY}
S.Hosono, A. Klemm, S. Theisen, S.-T. Yau, Comm. Math. Phys.
 {\bf 167} (1995) 301.

\item
\label{LY}
B.H.Lian, S.-T. Yau, hep-th/9507151, hep-th/9507153.

\item
\label{BK}
P. Berglund, S. Katz, Nucl. Phys. {\bf B420} (1994) 289.

\end{enumerate}

\end{document}